\documentclass[aps,pre,showpacs,amsmath,superscriptaddress,preprintnumbers,reprint]{revtex4-2}
\usepackage{graphicx,amsfonts,amssymb,multirow,array,esint, physics}
\usepackage{dcolumn}
\usepackage{bbold}
\usepackage{color,soul}
\usepackage[dvipsnames]{xcolor}
\definecolor{darkblue}{rgb}{0,0,0.6}
\usepackage[colorlinks,linkcolor=darkblue,citecolor=darkblue,urlcolor=darkblue]{hyperref}
\usepackage[utf8]{inputenc}
\usepackage{natbib}
\graphicspath{{Images/}}

\begin{document}

\title{Direct Boltzmann inversion method from particle configurations at arbitrary state points}

\author{Olivier Coquand}

\affiliation{Laboratoire de Modélisation Pluridisciplinaire et Simulations, Université de Perpignan Via Domitia, 52 avenue Paul Alduy, F-66860 Perpignan, France}

\author{Davide Paolino}

\affiliation{Gulliver, CNRS UMR 7083, ESPCI Paris, PSL Research University, 75005 Paris, France}

\author{Ludovic Berthier}

\email{ludovic.berthier@espci.fr}

\affiliation{Gulliver, CNRS UMR 7083, ESPCI Paris, PSL Research University, 75005 Paris, France}

\date{\today}

\begin{abstract}
We introduce a direct Boltzmann inversion method to infer the interaction potential in particle systems using as input particle configurations generated at an arbitrary state point of the system. Unlike iterative Boltzmann inversion, the proposed method does not require performing a new Monte Carlo simulation at each step of the iteration process. It relies instead on enforcing consistency between two independent estimates of the pair correlation function, respectively obtained from interparticle distances and from pairwise forces. As a result, the approach is computationally inexpensive and straightforward to implement. Because it relies on the sole expression of interparticle forces, our method naturally applies to any state point, including when the density is large and alternative methods may fail. Here we present the basic principles of the method and benchmark its performance on a diverse set of test potentials studied using computer simulations. Practical aspects and detailed implementation of the method are also discussed. Owing to its simplicity and generality, the method should be broadly applicable, from the construction of coarse-grained interaction potentials to the inference of effective interactions in non-equilibrium systems.
\end{abstract}

\maketitle

\section{Introduction}

In liquid-state theory~\cite{hansenTheorySimpleLiquids2013}, two cornerstone quantities are the interaction potential, $u(r)$, and the radial distribution function, $g(r)$. The first dictates the microscopic dynamics of each particle, while the second describes the structure. The latter can easily be obtained numerically, or experimentally using for example scattering or microscopy techniques. In a classic paper, Henderson proved that for equilibrium systems governed by pairwise interactions, there exists a one-to-one correspondence between the pair potential and the radial distribution function~\cite{hendersonUniquenessTheoremFluid1974}. A fundamental consequence is that knowledge of one function uniquely determines the other, at least in principle. As is well-known, there is no exact analytic formula to predict directly $g(r)$ from $u(r)$, and the derivation of approximate closure relations was a basic endeavor of liquid-state theory~\cite{hansenTheorySimpleLiquids2013}. The parallel development of computer simulations of particle systems~\cite{allenComputerSimulationLiquids2017,frenkel2023understanding} has allowed for an essentially exact determination of $g(r)$ given any interaction potential $u(r)$. From a numerical viewpoint, the forward problem of determining $g(r)$ for a given $u(r)$ can therefore be considered as being essentially solved.  

By contrast, the inverse problem~\cite{chayesInverseProblemClassical1984} of computing the interaction potential from a known $g(r)$ has attracted somewhat less attention, and there exists no fully satisfying solution either analytically~\cite{hansenTheorySimpleLiquids2013,chakrabarty2011high} or numerically. This so-called Boltzmann inversion problem is nevertheless of interest as it can be used for several purposes. An obvious application is the determination of an unknown interaction potential from an experimental dataset~\cite{royallMeasuringColloidalInteractions2007}. For pairwise interactions, this can be done by studying only two particles, but there is no such simple method available when an arbitrary bulk state point is accessible, or in the presence of many-body interactions. Another application is the determination of an effective coarse-grained interaction potential representing a more complex system with a larger number of degrees of freedom. This is a useful task, as the determined coarse-grained model can then be studied numerically over much larger length scales and time scales~\cite{maurelMultiscaleModelingPolymer2015,ingolfssonPowerCoarseGraining2014,mooreDerivationCoarsegrainedPotentials2014}. A related application is the deduction of optimal effective pair potentials for systems interacting with non-pairwise additive interactions. A final application could be the determination of effective interactions in systems evolving far from equilibrium~\cite{torquatoPreciseDeterminationPair2022,rees2026effective}, in order to provide an equilibrium interpretation of physics phenomena observed far from equilibrium. 

Over the years, a number of numerical methods have been developed to solve the inverse Boltzmann problem. Existing methods can be broadly divided into model-based and model-free approaches. Model-based methods essentially fit the measured radial distribution function by optimizing a parametrized family of potentials. Although widely used~\cite{torquatoPreciseDeterminationPair2022,wangEquilibriumStatesCorresponding2023,tianDeterminationPairwiseInteractions2023}, these approaches rely heavily on guessing an accurate parametrization and become less tractable as the number of parameters increases~\cite{izvekovEffectiveForceFields2004}. Model-free techniques, by contrast, do not impose a functional form for $u(r)$. These algorithms typically refine iteratively an initial potential guess until the simulated structural data matches the reference one~\cite{delbaryGeneralizedNewtonIteration2020}. The main differences between different algorithms lie in the iteration rules~\cite{bernhardtIterativeIntegralEquation2021}, whose complexity typically trades off with the number of iterations needed to converge. For example, Inverse Monte Carlo~\cite{lyubartsevCalculationEffectiveInteraction1995} converges in relatively few steps but this requires long simulations at each step for accurate statistics as well as a good initial guess~\cite{murtolaCoarsegrainedModelPhospholipid2007}. In contrast, Iterative Boltzmann Inversion~\cite{schommersPairPotentialLiquid1973,reithDerivingEffectiveMesoscale2003,mooreDerivationCoarsegrainedPotentials2014} relies on a very simple update rule, making it one of the most robust and widely-used algorithms, applied in fields from liquid interfaces~\cite{jochumStructurebasedCoarsegrainingLiquid2012} to polymers~\cite{maurelMultiscaleModelingPolymer2015} and biomolecules~\cite{ingolfssonPowerCoarseGraining2014}. Its convergence is typically much slower than Inverse Monte Carlo, often requiring an order of magnitude more iteration steps~\cite{jainInverseMonteCarlo2006}. Recent proposals using Machine Learning to reconstruct interactions are promising~\cite{berressem2021boltzmann}, but still lack the accuracy of iterative algorithms~\cite{ruiz-garciaDiscoveringDynamicLaws2024}, or must rely on intense training~\cite{bag2021interaction}.   

A common feature of all iterative methods mentioned above is the need to perform a computer simulation at each iteration step using the updated potential, in order to compare the generated data to the target $g(r)$. This represents the main computational bottleneck when performing the inversion, potentially limiting scalability to larger systems or higher accuracy.

Recently, an elegant strategy was proposed to bypass this bottleneck~\cite{stonesModelFreeMeasurementPair2019,rees-zimmermanInvertingGrUr2025,rees-zimmermanNumericalMethodsUnraveling2025,rees2026effective}. The method uses two expressions for the radial distribution function, the usual one using pairwise distances and a second one known as test-particle insertion formula~\cite{hansenTheorySimpleLiquids2013,stonesMeasuringManybodyDistribution2023} that requires knowledge of the interaction potential. The quality of the updated potential is now assessed by comparing the radial distribution function generated by the test-particle insertion to the reference one coming from the initial data set. An important shortcoming of this approach is that the test-particle insertion method is computationally much more expensive than conventional $g(r)$ calculations. An even more severe limitation is that particle insertions become virtually impossible in dense fluids and the proposed approach is thus by construction limited to small enough densities to allow particle insertions~\cite{rees-zimmermanInvertingGrUr2025}. 

Here, we propose an alternative approach to directly obtain the interaction potential without performing a novel computer simulation at each step of the iteration and that can be applied to any state point, including dense fluids. Our central idea is to replace the particle insertion expression of the radial distribution with an alternative expression based on interparticle  forces~\cite{borgisComputationPairDistribution2013,rotenbergUseForceReduced2020}. This force-formula for $g(r)$ involves both interparticle distances and forces. Computationally, it is just as cheap as the conventional histogram method, and it is applicable and accurate at arbitrary state points, with no limit on the studied density. Our work therefore considerably simplifies and extends the applicability of the recent efforts performed in Refs.~\cite{stonesModelFreeMeasurementPair2019,rees-zimmermanNumericalMethodsUnraveling2025} to directly solve numerically the Boltzmann inversion problem.  

Here, we explain how to implement numerically this direct inversion method to obtain the interaction potential from a given data set that consists of a series a particle configurations. The method converges rapidly, accurately, and is computationally inexpensive. We implement and demonstrate its efficiency in a variety of situations for which the interaction potential is known, and we show that our inversion method accurately recovers the correct potential in all cases in just a few minutes, thus efficiently solving the Boltzmann inversion problem. We leave applications to experimental data, coarse-graining and non-equilibrium situations for a future work. 

The manuscript is organised as follows. 
The central idea behind the direct inversion method is presented in Sec.~\ref{sec:algorithm}.
The practical implementation of the algorithm is discussed in Sec.~\ref{sec:practical}. 
We implement, test and validate the algorithm on a number of known potentials at various state points in Sec.~\ref{sec:validation}. 
We discuss the results and present some outlook for future work in Sec.~\ref{sec:outlook}.

\section{Direct Boltzmann inversion algorithm}

\label{sec:algorithm}

\subsection{Radial distribution function (RDF)}

A many-body system of $N$ particles can be completely characterized by the knowledge of all the $n$-particle distribution functions $g_N^{(n)} (\vb* r_1, ..., \vb* r_n)$~\cite{hansenTheorySimpleLiquids2013}. The computation of all these quantities for $N>2$ is often difficult, but fortunately the knowledge of low-order particle distribution functions $n\leq2$ is often sufficient to evaluate the equation of state and other thermodynamic properties. In particular, if the system is both isotropic and homogeneous, the pair distribution function $g_N^{(2)} (\vb* r_1, \vb*r_2)$ only depends on the separation $r_{12} =  | \vb*r_2 - \vb*r_1 | $. In this case,  it is usually called radial distribution function (RDF) and referred as $g(r)$, taking the form
\begin{equation}
\label{eq: gr}
    g(r) = \expval{\frac{1}{\rho N} \sum_{i=1}^N \sum_{j \neq i} \delta (r-r_{ij})},
\end{equation}
where $r_{ij} =  | \vb*r_i - \vb*r_j |$, $\delta(x)$ is the Dirac delta function, $\rho = N/V$ the number density, and $\expval{\cdots}$ represents an ensemble average. 

\subsection{Conventional RDF estimate from pairwise distances}

\label{sec:histo}

Given the configurations of a homogeneous system, the radial distribution function $g(r)$ can be computed using a distance histogram (DH) procedure, that amounts to a simple discretization of Eq.~(\ref{eq: gr}). In practice, one counts the number of particles $N(r)$ within a spherical shell of radius $r$ and thickness $\Delta r$, normalized by the shell volume $\Omega_d r^{d-1} \Delta r$ and the particle density $\rho$, where $\Omega_d$ is the solid angle in $d$ dimensions ($\Omega_2 = 2\pi$ and $\Omega_3 = 4\pi$). Averaging over an ensemble of configurations $\expval{\cdots}$ yields
\begin{equation}
\label{eq: gr_histo}
     g_{DH}(r) = \frac{\expval{N(r)}}{\rho \Omega_d r^{d-1} \Delta r} .
\end{equation}
Note that Eq.~(\ref{eq: gr_histo}) depends explicitly on the choice of the bin size $\Delta r$, and the variance of the RDF obtained via this method diverges as $1/\Delta r^2$ for a given amount of data. The number of operations needed to measure $g_{DH}(r)$ scales quadratically with the number of particles, ${\cal O}(N^2)$ (this is a sum over particle pairs), and it assumes no prior knowledge of the interparticle potential. Nevertheless, Henderson's theorem implies that the RDF $g(r)$ evaluated via interparticle distances is uniquely related to the potential $u(r)$ used to generate the configurations for pairwise interactions~\cite{hendersonUniquenessTheoremFluid1974}. 

\subsection{Computing the RDF from interparticle forces}

\label{sec:gforce}

Inspired by previous work on quantum electronic densities, Borgis and coworkers proposed a new formula to estimate the pair correlation function of a homogeneous and isotropic system when the forces acting on each particle are known~\cite{borgisComputationPairDistribution2013}:
\begin{equation}
\label{eq: Borgis_out}
    g_{\text{force}}(r) = 1 - \frac{1}{\rho N }  \expval{
    \sum_{i<j} \beta( \vb f_i - \vb f_j )  \cdot  \frac{\vb* r_{ij}}{\Omega_dr_{ij}^d}     \Theta(r_{ij}-r) },
\end{equation}
where $\Theta(x)$ is the Heaviside function and $\beta=1/k_BT$, with $k_B$ the Boltzmann constant. In this expression, $\vb f_i$ represents the total force acting on particle $i$. This formula can be qualitatively understood as resulting from an integration by part of the original expression in Eq.~(\ref{eq: gr}), transforming the delta function into a Heaviside one, while the derivative on the Boltzmann distribution produces the force $\vb f_i$. While this formula was originally obtained using precisely this integration by part~\cite{borgisComputationPairDistribution2013}, we provide an alternative derivation in Appendix~\ref{appendix:RDF} that first derives (producing the forces) and then re-integrates (producing the Heaviside functions) the function $g(r)$ with respect to the variable $r$. This derivation is close to the one given in Ref.~\cite{frenkel2023understanding}. All derivations are of course mathematically equivalent.     

In particular, the derivations make it clear that no assumption is made on the nature of the potential $u(r)$ (for instance it is not necessarily pairwise additive). The formula in Eq.~(\ref{eq: Borgis_out}) differs from the original work~\cite{borgisComputationPairDistribution2013} by a factor $2$, as already corrected in Ref.~\cite{purohitForcesamplingMethodsDensity2019}. An important assumption is that the system is at thermal equilibrium so that the ensemble average is represented by an integration of the configuration space, weighted by the Boltzmann distribution. This formula can therefore not be applied to a non-equilibrium driven steady state, for instance.   

An important difference with respect to the histogram method is that a given pair of particles, $(i, j)$, contributes at all distances $r \leq r_{ij}$ instead of contributing to a single bin at distance $r=r_{ij}$. This helps reducing the variance of the $g(r)$ estimate~\cite{colesReducedVarianceAnalysis2021}, which no longer depends on the spatial discretization $\Delta r$ chosen to represent the RDF. As a result, the radial distribution function can be evaluated with an arbitrary spatial resolution without increasing the statistical noise.

It is important for the inversion method discussed in this work to realize that the evaluation of $g_{\text{force}}(r)$ at $r$ depends on all distances and is therefore no more local in position space. This unfortunately implies that a small error on the reconstructed potential $u(r)$ at some given $r$ may affect the function $g_{\text{force}}(r)$ at other $r$ values.

Another notable difference is that the function $g_{\text{force}}(r)$ obtained in Eq.~(\ref{eq: Borgis_out}) properly converges, by construction, to 1 in the limit $r \to \infty$, but displays a spurious non-zero value in the $r \to 0$ limit, which only vanishes in the limit of an infinite data set~\cite{colesReducedVarianceAnalysis2021}.  
These two differences will play an important role in the practical implementation discussed below. However, the key idea is that both $g_{\text{force}}(r)$ and $g_{\text{DH}}(r)$ provide unbiased estimates of the same analytical quantity $g(r)$, and both estimates must lead the same result $g(r)$ in the limit of infinite statistics. The coincidence of both estimates however only happens when the correct expression of the forces is employed in Eq.~(\ref{eq: Borgis_out}). Enforcing this condition will allow us to directly determine forces from an ensemble of particle configurations.   

\begin{figure}
\includegraphics[width=0.8\linewidth]{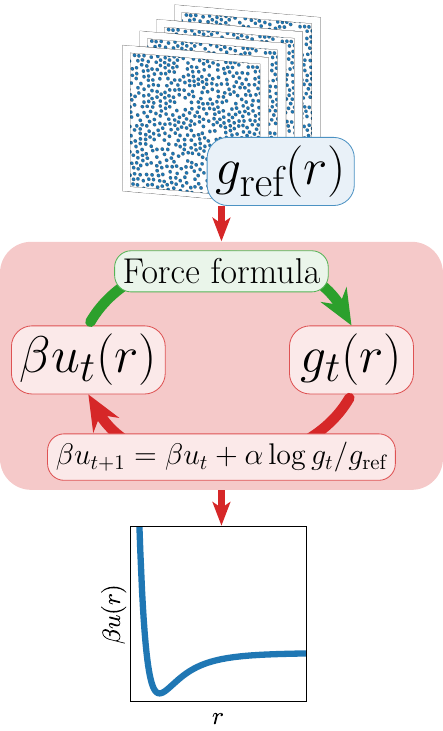}
\caption{Sketch of the direct Boltzmann inversion method. The available data set (series of images) produces a reference RDF $g_{\rm ref}(r)$. A guess potential $u_t(r)$ is used to generate the RDF $g_t(r)$ using the force formula, and the difference between $g_{\rm ref}(r)$ and $g_t(r)$ is used to make an improved guess for the potential $u_{t+1}(r)$. The fixed point provides the correct inverted potential $u(r)$.} 
    \label{fig:schematic}  
\end{figure}

\subsection{The direct inversion algorithm}

We are now in a position to explain our direct inversion algorithm, which we have sketched in Fig.~\ref{fig:schematic}.

It is useful to recall first the traditional Iterative Boltzmann Inversion method. This starts by the determination of a reference RDF, $g_{\rm ref}(r)$, obtained from the data set using the distance histogram method. The iteration loop can then start. An initial guess of the interaction potential, $u_0(r)$, is used in a Monte Carlo simulation to determine the RDF $g_0(r)$ using the histogram method. The difference between $g_0$ and $g_{\rm ref}(r)$ is used to make an improved guess for the potential, $u_1(r)$, leading to $g_1(r)$. The process is then repeated until the guess $u_t(r)$ made at step $t$ leads to a measured $g_t(r)$ that is equal to $g_{\rm ref}$. At this point, the algorithm has reached a fixed point, and the potential $u_t(r)$ is no longer updated.  

Our approach follows the same philosophy and starts with the construction of the reference RDF, $g_{\rm ref}(r)$, and a guess $u_0(r)$ is again made for the potential. Crucially, we now obtain the RDF $g_0(r)$ from the guess $u_0(r)$ using the force formula applied on the original data set. (A very similar idea was proposed before by Stones and coworkers~\cite{stonesModelFreeMeasurementPair2019}, but they employed the test-particle insertion estimate of the RDF instead of the force formula.) The difference between $g_0$ and $g_{\rm ref}(r)$ is used to make an improved guess for the potential, $u_1(r)$. The process is then repeated until convergence. In contrast to Iterative Boltzmann Inversion, the only data set used in the iteration is the original one, as the process does not require generation of new data via Monte Carlo simulations.   

In all inversion schemes, an improved guess of the potential from the measured difference between two RDF estimates needs to be made. Here, we follow the update rule proposed by Schommers~\cite{schommersPairPotentialLiquid1973}: 
\begin{equation}
\label{eq: IBI-Schommers scheme}
    \beta u_{t+1}(r) = \beta u_t(r) + \alpha \log \left( \frac{g_t(r)}{g_{\text{ref}}(r)} \right) , 
\end{equation}
where $\alpha \in [0, 1]$ is an empirical regularization factor used to control the stability of the iteration procedure. By construction, the potential $u_t(r)$ is not updated when the RDF at step $t$ agrees with the target reference. 

This update formula remains however largely empirical. For consistency, we summarize Schommers' justification of Eq.~(\ref{eq: IBI-Schommers scheme}) in Appendix~\ref{Apa}. It has the desired property that the fixed point $(g^*(r), u^*(r))$ necessarily corresponds to $u^*(r)=u(r)$ and $g^*(r)=g_{\rm ref}(r)$. We discuss the stability of this fixed point in Appendix~\ref{Apb}. 

\section{Practical implementation}

\label{sec:practical}

While the workflow presented in Fig.~\ref{fig:schematic} faithfully describes all steps of the direct Boltzmann inversion algorithm, there are several practical issues that arise during its implementation, which we now discuss.  

\subsection{Generation of reference RDF from data}

\label{sec:gref}

The reference radial distribution function $g_{\text{ref}}(r)$ is constructed from the microscopic configurations available in the original data set. At this point, the 
distance-histogram method in Eq.~(\ref{eq: gr_histo}) needs to be used, since it does not require knowledge of the pair potential. As discussed in Sec.~\ref{sec:histo}, the formula contains an adjustable parameter $\Delta r$ used to bin the pair distribution, that requires a careful discussion.

Ideally, one would like $\Delta r$ to be as small as possible in order to reconstruct an interaction potential $u(r)$ on a sufficiently fine grid to allow the numerical determination of the derivative $u'(r)$ involved in force estimates. However, the distance-histogram RDF becomes noisy when $\Delta r$ is chosen too small for the amount of available data. 

\begin{figure}[t]
    \includegraphics[width=1\linewidth]{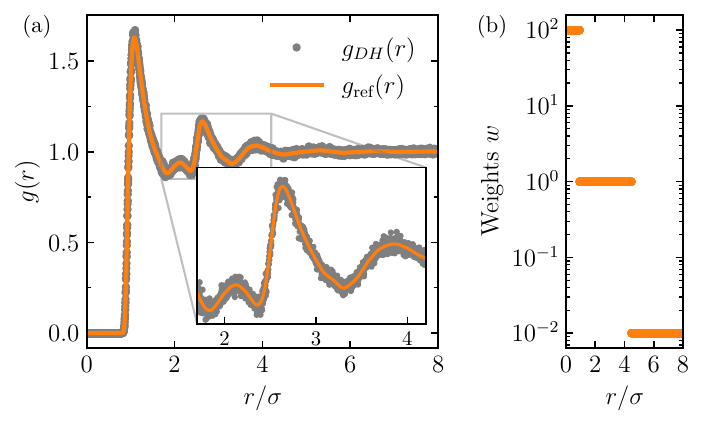}
    \caption{(a) The RDF measured by the distance-histogram $g_{DH}(r)$ is smoothed and finely discretized to produce the reference $g_{\rm ref}(r)$. (b) Weights used for the spline interpolation. Data taken for the shoulder potential at $\rho\sigma^2=0.56$ and $\beta \epsilon =0.5$.}
    \label{fig:rdf_smoothing}
\end{figure}

To mitigate these effects, we construct $g_{\rm ref}(r)$ in two steps. We first use the conventional method that decreases $\Delta r$ until the RDF becomes noisy. We then smooth and interpolate the measured function to finally obtain a smooth, finely discretized reference RDF $g_{\rm ref}(r)$. The smoothing and interpolation procedures are illustrated in Fig.~\ref{fig:rdf_smoothing}. 

In practice, we apply spline interpolation to the measured RDF. This reduces noise by fitting the histogram data with a continuous function that can then be evaluated at any desired resolution. Given a dataset $\mathcal D = \{(r_i, g_i)\}_{i=1}^M$, the spline is obtained by minimizing the following loss functional
\begin{equation}
\label{eq: 3_spline}
    \mathcal L[f] = \sum_{i=1}^M w_i\abs{g_i - f(r_i)  }^2 + \lambda \int \abs{\pdv[2]{f}{r}}^2\dd r ,
\end{equation}
using piecewise cubic polynomials for $f(r)$. The first term measures deviation between the fit function and the data points, while the second penalizes curvature, enforcing smoothness of the interpolated function $f(r)$. Here, $\lambda$ tunes the trade-off between data fidelity (first term) and smoothness (second term).   

While conceptually easy, applying a single smoothing function across the entire radial distribution function is challenging because the RDF behaves very differently across different scales. At short distances, the hard-core region is undersampled and exhibits a sharp rise from zero, that splines may fail to capture. At larger range, the RDF peaks may suffer from statistical noise in the distance-histogram method, and this may obscure fine structural details. Finally, at long distances where $g(r) \approx 1$, noise dominates, making the interpolation potentially unreliable due to overfitting.

To address these issues, we implement a region-specific weighting scheme $w_i$ with three distinct zones: the hard-core region (from $r=0$ to the midpoint to the first peak) receives a weight $10^2$ times stronger than the intermediate region with unit weight. Instead, the noisy long-range region is deliberately suppressed with a weight reduced to $10^{-2}$. Lastly, the smoothing parameter $\lambda$ is determined automatically by the \texttt{scipi.signal} library~\cite{virtanenSciPy10Fundamental2020} through generalized cross-validation. This approach maintains simplicity and transferability across diverse RDF shapes while delivering robust performances for the inversion procedure. As shown in Fig.~\ref{fig:rdf_smoothing}, the obtained reference $g_{\rm ref}(r)$ is simultaneously faithful to the measured data, smooth, and can be discretized over an arbitrary fine spatial grid. 

\subsection{Choice of cutoff distances}

Our goal is to reconstruct a function $u(r)$ finely discretized over a range of distances. While in principle we would like that range to be between $r=0$ and $r=\infty$, in practice one needs to restrict this idealized range to a finite one and introduce two cutoff values to reconstruct $u(r)$ over a finite range $r \in [r_{\rm low}, r_{\rm cut}]$. 

An upper cutoff $r_{\rm cut}$ is needed when forces and potentials become very small at large distances. In that case, there will not be enough signal in the measured data set to reconstruct very small numbers. In a typical application where that cutoff is not known, the chosen value could be refined iteratively by starting from a high value and progressively lowering $r_{\rm cut}$ while monitoring stability and convergence of the inversion algorithm. 

For strongly repulsive potentials, a lower bound $r_{\text{low}}$ is also needed. Since the hard core part of the potential is not explored by the particles, it is obvious that the data set contains no information about the functional form of $u(r)$ at very short distances. On the other hand, we also find that taking $r_{\text{low}}$ too large can lead to an incorrect determination of the potential $u(r)$. The reason is related to the discussion in Sec.~\ref{sec:gforce} above. When $r_{\text{low}}$ is too large, we miss pairs of particles in the data set for which $r_{ij} < r_{\rm low}$ and this may affect the entire function $g_{\rm force}(r)$. In turn, an incorrect estimate of $g_{\rm force}(r)$ will lead to an incorrect fixed point of the algorithm. 

In practice, we go over the data set and measure over independent configurations the minimum distance found in each configuration. We then set $r_{\text{low}}$ as the average of this quantity. This ensures that very few particle pairs are missed and that we do not attempt to reconstruct the potential in a region where statistics is poor. 

Particle pairs with separations $r_{ij} < r_{\text{low}}$ are excluded by the Heaviside function in the calculation of $g_\text{force}(r)$ in Eq.~(\ref{eq: Borgis_out}). However they still contribute significantly to the total forces $\mathbf{f}_i$ and $\mathbf{f}_j$. To account for these contributions, the derivative of the potential (that is needed for the force calculations) in the region $r < r_{\text{low}}$ is extrapolated as:
\begin{equation}
    u'(r) = u'(r_{\text{low}}) \left( \frac{r_\text{low}} {r} \right)^2.
\label{eq:extrapolation}
\end{equation}
This expression ensures continuity of the force at $r=r_\text{low}$, and effectively introduces a soft-core repulsion that prevents nonphysical artifacts during the inversion process. We checked that the specific choice made in Eq.~(\ref{eq:extrapolation}) is arbitrary and largely irrelevant, as several other choices can also be employed. 

\subsection{Initialization of the potential}

To run the iterative inversion scheme one should finally choose an initial potential guess, $\beta u_0(r)$. While the final result should be independent of this initial choice, it is obvious that starting from a totally incorrect guess may lead to a much longer iterative process, or even introduce numerical instabilities. Moreover, the Schommers scheme in Eq.~(\ref{eq: IBI-Schommers scheme}) assumes that the initial potential guess is a sufficiently good approximation of the unknown pair potential~\cite{schommersPairPotentialLiquid1973}. For these reasons, typically the starting potential is chosen as 
\begin{equation}
    \beta u_0 (r) = - \ln g_{\text{ref}}(r)\, , 
\end{equation}
which is also called potential of mean-force. This guess coincides with the true potential in the case of weak interactions. This choice provides a model-agnostic initial guess which moreover reproduces the most salient features of the true potential, thus being close enough in the sense of Schommers (see appendix~\ref{Apa}). We adopt this choice in the following.

\subsection{Improved iteration formula to preserve physical constraints}

Since the force-based estimator in Eq.~(\ref{eq: Borgis_out}) can yield spurious finite values as $r\to0$~\cite{colesReducedVarianceAnalysis2021}, the estimate  $g_t(r)$ may in particular become negative during the inversion process, especially when the initial potential guess is poor. To prevent logarithmic singularities in the Schommers update rule~(\ref{eq: IBI-Schommers scheme}) in that case, we apply a vertical shift to $g_t(r)$ to ensure the argument of the logarithm always remains positive.

Specifically, we define  $r_\text{min}$ as the radial distance where the current estimate $g_t(r)$ reaches its minimum over the inversion window $[r_{\rm low}, r_{\rm cut}]$. We then implement the modified update rule:
\begin{equation}
\beta u_{t+1} = \beta u_t + \alpha \log \left( \frac{g_t(r) - \delta g}{g_\text{ref}(r)} \right),
\label{eq:updateshift}
\end{equation}
where the offset $\delta g \equiv g_t(r_\text{min}) - g_\text{ref}(r_\text{min})$ ensures that the numerator is well-behaved and matches the positive reference $g_\text{ref}(r)$ at its minimum value. In practice, $r_\text{min}$ typically coincides with $r_\text{low}$, the lower bound of the inversion window, as expected. Instances where $g_t(r)$ becomes negative at larger radial distances are rare and are generally restricted to the earliest iterations steps. As the potential converges toward a fixed point, we checked that $\delta g$ also becomes very small, and Eq.~(\ref{eq:updateshift}) becomes in that case mathematically equivalent to the conventional Schommers expression, in particular near the fixed point.

\subsection{Convergence metrics}

Finally, an iterative algorithm requires a criterion to determine when convergence has been reached. Given some distance metric $D(f, g)$ between two functions $f(x)$ and $g(x)$ defined on a discretized series of $x$ values, one could theoretically stop at step $T$ when the current estimate $g_{T}(r)$ is closer to the target than a desired precision $\nu$, namely
\begin{equation}
\label{eq: 3_convergence with reference}
     D(g_{T} , g_{\text{ref}}) \leq \nu . 
\end{equation}
However, we specifically adopt a convergence criterion based on the distance between successive iterations:
\begin{equation}
    D(g_{T} , g_{T-1}) \leq \nu.
\end{equation}
This second definition has several advantages: it is more easily transferable across different pair distribution shapes and does not assume that the distance to the reference decreases monotonically.

Because our method is computationally inexpensive, we have set a stringent precision requirement of $\nu=10^{-10}$ to stop the iteration loop. This ensures the solution has reached the fixed point. This very low value eliminates any residual error that might exist when more costly methods are used and the iteration is stopped early.

In practice, we verified that both criteria yield nearly identical results, confirming the robustness of the fixed-point approach. The distance $D(f,g)$ is chosen as the mean-squared error
\begin{equation}
    D(f, g) = \frac{1}{n} \sum_{i=1}^{n} \abs{f(r_i) - g(r_i)}^2,
\end{equation}
computed over the window $[r_{\text{low}}, r_{\text{cut}}]$ discretized in $n$ bins.

\section{Validation using known potentials}

\label{sec:validation}

The ultimate goal of our inversion algorithm is to reconstruct the pair potential $u(r)$ starting from equilibrium configurations of the system. As a first test, we use configurations obtained through numerical simulations of known potentials, in order to assess the quality of the inversion method and detect possible pitfalls.

\subsection{Four simulated potentials}

\label{sec:potentials}

We analyze different two-dimensional systems of monodisperse particles of mass $m$, interacting with various widely used potentials, each one posing different challenges for their inversion. The first one is the conventional Lennard-Jones (LJ) potential~\cite{lennard1931cohesion}, truncated and shifted at $r_{c}^{\text{LJ}} = 2.5\sigma$:
\begin{equation}
    u_{\text{LJ}}(r)  = 
    \begin{cases}
        4\epsilon \qty[\qty(\frac{\sigma}{r})^{12} - \qty(\frac{\sigma}{r})^{6} + C] & r\leq r_c^{\text{LJ}},\\
        0 &\qq*{otherwise},
    \end{cases}
\end{equation}
where $C$ is an additive constant so that $\beta u_{\text{LJ}}(r_c^{\text{LJ}}) = 0$. The Lennard-Jones potential presents both a repulsive soft core and an attractive long-range tail. 

The second tested potential is the Weeks-Chandler-Anderson (WCA) potential~\cite{weeksRoleRepulsiveForces1971}, which is nothing but the LJ potential truncated at its minimum $r_{c}^{\text{WCA}} = 2^{1/6} \sigma$:
\begin{equation}
    u_{\text{WCA}}(r)  = 
    \begin{cases}
        4\epsilon \qty[\qty(\frac{\sigma}{r})^{12} - \qty(\frac{\sigma}{r})^{6} ] + C^\prime,  & r < r_c^{\text{WCA}}, \\
        0 &\text{otherwise},
    \end{cases}
\end{equation}
where $C^{\prime}=\epsilon$ is such that $u_{\text{WCA}}(r_c) = 0$, thus creating a purely repulsive potential that is often used in high-density regimes where the role of attractive forces diminishes.

\begin{figure*}
    \includegraphics[width=\linewidth]{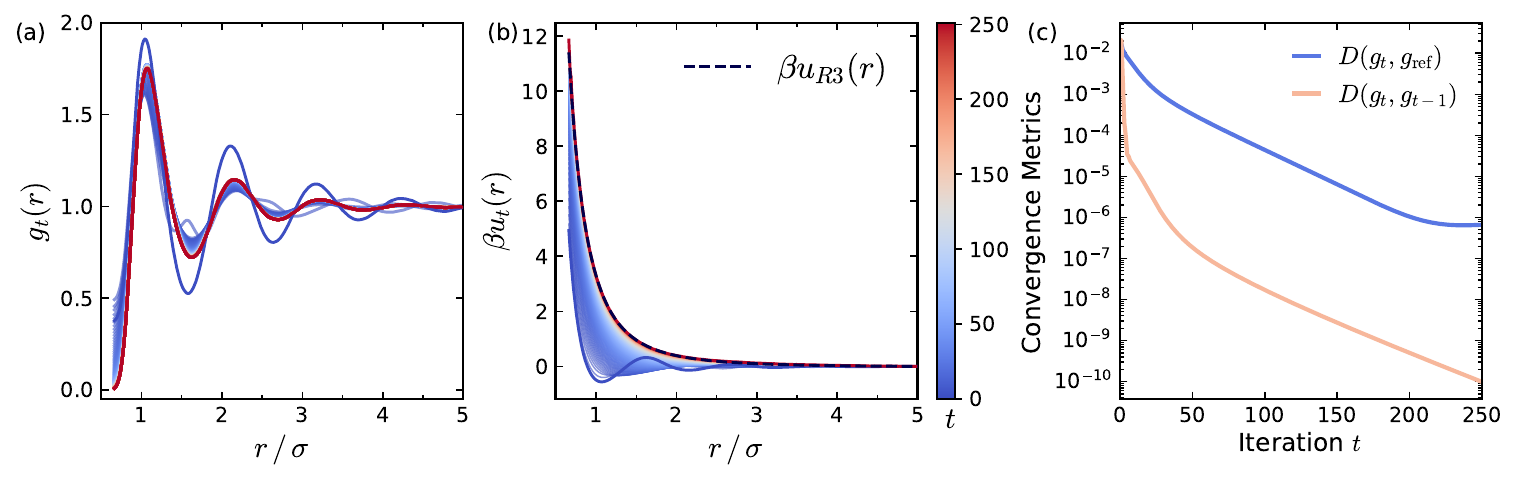}
    \caption{Convergence of the iterative procedure for an inverse cubic potential at $\rho\sigma^2 = 0.80$ and $\beta \epsilon = 10/3$, using $\alpha=0.5$. (a) Evolution of $g_t(r)$ toward the reference (red line). (b) Effective potentials $\beta u_t(r)$ reconstructed on the window $[0.663\sigma, 5\sigma]$; dashed line shows the analytical potential and the thick blue line indicates the initial guess $\beta u_0 = -\log g_{\text{ref}}(r)$. (c) Convergence metrics as a function of the iteration number $t$.}
    \label{fig: 3}
\end{figure*}

Thirdly, we introduce a long-range power-law potential that scales as the inverse of the cubic distance:
\begin{equation}
    u_{R3}(r) = 
    \begin{cases}
    \epsilon \qty(\frac{\sigma}{r})^3 + C^{\prime\prime}& r\leq r_c^{{R3}}= 5\sigma,\\
     0 & \text{otherwise}, 
    \end{cases}
\end{equation}
where again $C^{\prime\prime}$ is an additive constant to ensure $u_{R3}(r_c^{R3})=0$. Such $r^{-3}$ potential describes softer repulsive interactions compared to LJ and WCA, with slower decay relevant for dipole-like or screened electrostatic effects, common in soft colloidal systems~\cite{zahnTwoStageMeltingParamagnetic1999}.

Lastly, we introduce a shoulder potential~\cite{gribovaWaterlikeThermodynamicAnomalies2009}
\begin{equation}
    u_{\text{SH}}(r) = 
    \begin{cases}
        \epsilon\qty(\frac{\sigma}{r})^n - \frac{\epsilon}{2} \tanh\frac{k_0}{\sigma}(r - r_0) + C''', & r < r_c^{\text{SH}} \\
        0, & \text{otherwise}
    \end{cases}
\end{equation}
that features two characteristic length scales. It has a hard core $r^{-n}$, and an outer softer shell represented by the hyperbolic tangent. The parameters are fixed as in Ref.~\cite{rees-zimmermanNumericalMethodsUnraveling2025}, namely $n=14$, $k_0 =10$ and $r_0 = 2.5\sigma$ with a cut-off at $r_{c}^{\text{SH}} = 2.8\sigma$. Once more, the potential is shifted to vanish at its cutoff, fixing the value of the constant $C'''$.

We consider square systems in two dimensions with periodic boundary conditions and box size $L = 60 \sigma$, while the number of particles $N$ and the temperature $T$ vary between different simulations. The values for $\rho = N/L^2$, the temperature $T$, and other simulation parameters will be reported in the figure captions accompanying the inversion results.

In practice, we first equilibrate and then sample equilibrium configurations in the canonical ensemble via molecular dynamics simulations using the open-source software LAMMPS~\cite{thompsonLAMMPSFlexibleSimulation2022}. Equations of motions are integrated using the Verlet algorithm, and the temperature is fixed via the Nosé-Hoover thermostat~\cite{allenComputerSimulationLiquids2017}. For each potential, the target radial distribution function $g_{\text{ref}}(r)$ is computed from 500 independent configurations using the distance-histogram method. 

\subsection{Inversion algorithm at work}

Having obtained the reference RDF $g_{\rm ref}(r)$ as explained in Sec.~\ref{sec:gref}, we now enter the iteration loop. In this part, we use 125 independent configurations produced by the molecular dynamics simulations.

We illustrate the results in Fig.~\ref{fig: 3} for the inverse cubic potential $u_{R3}(r)$. 
We use $\alpha = 0.5$, and reconstruct the potential over the range $[0.663\sigma, 5\sigma]$, using the potential of mean force as the initial guess, $\beta u_0(r) = - \log g_{\rm ref}(r)$. 
We show the evolution of both the radial distribution function $g_t(r)$ (Fig.~\ref{fig: 3}a) and the reconstructed potential $\beta u_t(r)$ (Fig.~\ref{fig: 3}b) as the iteration progresses. Convergence is monitored using both $D(g_t,g_{\rm ref})$ and $D(g_t, g_{t+1})$, see Fig.~\ref{fig: 3}c. 

From the evolution of the potential, we observe that the nonphysical double-well present in the initial guess is rapidly lost after a few iteration steps. Nonetheless, matching the precise values and functional form of the reference potential requires many additional iterations. 

Convergence metrics show that both the difference between successive iterations and the difference relative to the target drop below ten parts per million after roughly 160 iterations. Notably, the former drops by a further two orders of magnitude in twenty steps, signaling the arrest of the iteration procedure and the approach to the fixed point. 

In terms of computational performance, the entire procedure for this system requires about ten minutes on a standard laptop. Each step of the iteration simply requires the calculation of the radial distribution function over 125 configurations composed of $N=2916$ particles each. This could even be shortened if a less precise inverted potential were needed, but this method therefore represents a computationally very cheap inversion method. To achieve a similar performance using Iterative Boltzmann inversion would have required to perform 200 long and independent Monte Carlo simulations at each iteration step.  

\subsection{Results for other potentials}

We now collect our results for the different interaction potentials introduced in Sec.~\ref{sec:potentials}. In order to ease the discussion, we set the inversion cutoff $r_{\text{cut}}$ to match the simulation cutoff $r_c$ for each potential. We checked that increasing $r_{\text{cut}}$ beyond $r_c$ does not affect the quality of the inversion, with converged potential curves that coincide well with the ones presented here. 

\begin{figure*}
    \centering
    \includegraphics[width=\textwidth]{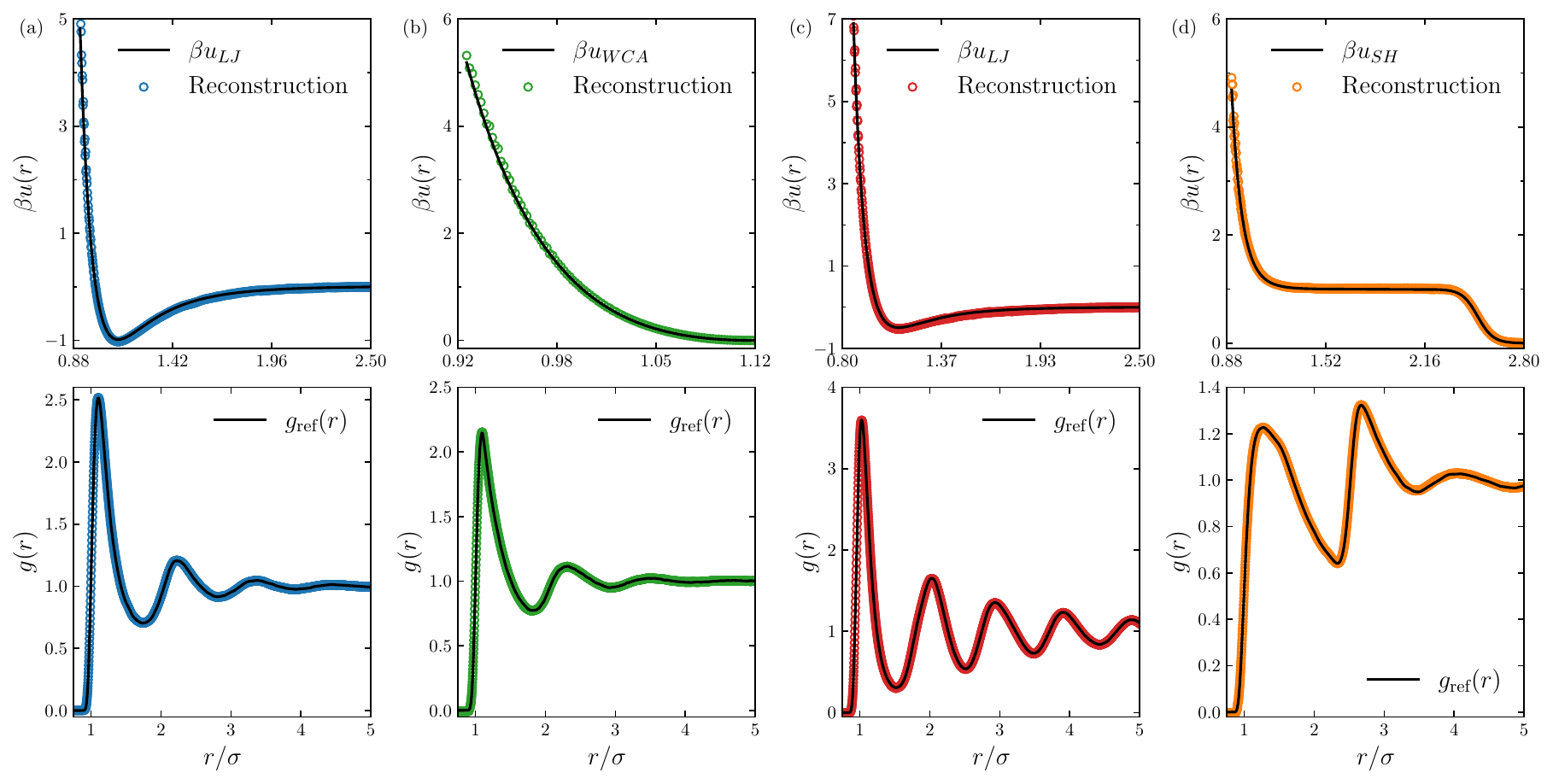}
    \caption{Potential reconstruction results for various interaction types. The top row displays the reconstructed pair potentials $\beta u(r)$, while the bottom row shows the corresponding radial distribution functions $g(r)$. In all panels, the solid black lines represent the reference and the colored scatter points denote the final reconstruction results 
    (a) Lennard-Jones (LJ) at $\rho\sigma^2 = 0.56, \beta\epsilon = 1.0, \alpha = 0.4$, reconstructed over the window $[0.915\sigma, 2.5\sigma]$; 
    (b) Weeks-Chandler-Andersen (WCA) at $\rho\sigma^2 = 0.56, \beta\epsilon = 1.0, \alpha = 0.2$, reconstruction window $[0.920\sigma, 2^{1/6}\sigma]$; 
    (c) LJ at $\rho\sigma^2 = 0.92, \beta\epsilon = 1/2, \alpha = 0.5$, reconstruction window $[0.863\sigma, 2.5\sigma]$; 
    (d) Shoulder potential at high density $\rho\sigma^2 = 0.28, \beta\epsilon = 1.0, \alpha = 0.2$, reconstruction window $[0.911\sigma, 2.8\sigma]$.}
    \label{fig:potentials}
\end{figure*}

The long-range $r^{-3}$ potential, often problematic for inverse methods due to its slow decay, is also reconstructed, as already shown in Fig.~\ref{fig: 3}.

All the other results are shown in Fig.~\ref{fig:potentials}. The overall agreement between the reconstructed and target potentials is excellent, even for particularly challenging systems. 

Both the Lennard-Jones potential (Fig.~\hyperref[fig:potentials]{4a}), characterized by steep short-range repulsion and a shallow attractive tail, and the WCA potential (Fig.~\hyperref[fig:potentials]{4b}), which is non-zero over a very narrow region, are accurately resolved, highlighting the robustness of the method in handling both attractive regimes and short-range interactions.

Remarkably, the method remains robust even at high densities ($\rho\sigma^2=0.92$,  see Fig.~\ref{fig:potentials}c), accurately recovering the potential in a state point where the test-particle insertion method would typically encounter severe convergence issues~\cite{rees-zimmermanInvertingGrUr2025}.

Finally, the shoulder potential (Fig.~\ref{fig:potentials}d), featuring both a steep repulsive core and an intermediate plateau, is recovered, demonstrating the ability of the method to capture potentials with multiple intrinsic length scales.

Across all test cases, potential reconstructions exceed expectations, with discrepancies appearing primarily in the derived force profiles. However, this limitation comes from the numerical differentiation used to compute forces. In applications where the goal is force reconstruction, additional post-processing, such as smoothing or filtering of the underlying potential, could significantly improve the quality of the extracted force field, if needed. 

\section{Summary and outlook}

\label{sec:outlook}

In conclusion, we have introduced a simple, robust, computationally cheap, and flexible Boltzmann inversion algorithm capable of reconstructing pair potentials directly from an ensemble of configurations obtained at arbitrary state points.

The method exploits the robustness of iterative Boltzmann inversion but a substantial reduction in computational cost is achieved by replacing expensive Monte Carlo simulations at each iteration step with a simple evaluation of the RDF using the force formula over the initial data set. This modification enables hundreds of iterations within seconds or minutes, which provides superior accuracy compared to traditional inversion schemes. The benchmark tests contained in our manuscript demonstrate that the method successfully recovers interaction potentials across a wide variety of physical situations, including steeply repulsive cores, short-ranged interactions, and potentials with multiple or very large characteristic length scales. 

Beyond its technical success, the approach holds significant promise for practical applications. Its efficiency makes it well-suited for analyzing experimental data, where structural measurements are often available but the underlying interactions may remain elusive. Moreover, it provides a viable tool for probing non-equilibrium systems through effective equilibrium mappings, and for developing coarse-grained models in soft matter and biomolecular contexts. In the latter case, an explicit derivation starting from the full many-body Hamiltonian would be requested. For these systems, our method would be particularly valuable, since the coarse-graining could easily be performed at any state point, and the evolution of effective interactions with thermodynamic parameters could thus be followed explicitly~\cite{louis2002beware}. All these directions will be the focus of future work, where we plan to apply the algorithm to experimental systems, non-equilibrium active matter, and coarse-graining problems.

\section*{Data Availability}

The data that support the findings of this study are openly available on Zenodo~\cite{zenodo_inversion}.

\acknowledgments

We thank R. Jack for useful discussions, D. Aarts and C. Valeriani for exchanges over their related work, F. van Wijland for discussions related to stability of the algorithm, and D. Frenkel and B. Guiselin for their interest in the force formula. We thank Yuan Liu, Qiuju Chen, Xurui Li, and Jianxiang Tian for work done in the early stages of this project~\cite{tian}. We acknowledge the financial support of the French Agence Nationale de la Recherche (ANR), under grants ANR-20-CE30-0031 (project THEMA) and ANR-24-CE30-0442 (project GLASSGO).

\bibliography{Inversion}

\clearpage

\appendix 

\section{Another derivation of the force formula}

\label{appendix:RDF}

The radial distribution function (RDF) provides a statistical description of fluid structure. Its definition is:
\begin{equation}
g(\vb*r) = \frac{1}{\rho N} \expval{\sum_{i\neq j} \delta(\vb*r-\vb*r_{ij})},
\qquad \vb*r_{ij}=\vb*r_i-\vb*r_j,
\end{equation}
where $\rho$ is the number density, $N$ is the number of particles. The configurational average is taken over the Boltzmann distribution for a system with potential energy $U = U(\{\vb*r_k\}, k=1 \cdots N)$:
\begin{align}
\langle\cdots\rangle & = \frac{1}{Z} \int \mathrm \dd^N \vb*r \ (\cdots)\, e^{-\beta U},\\
Z &=\int \mathrm \dd^N  \vb*r \ e^{-\beta U} ,
\end{align}
where $\beta=1/k_BT$.

We compute the radial derivative $ \partial_r g(\vb*r)$ by differentiating inside the expectation value:
\begin{equation}
\partial_r g(\vb*r) =  \frac{1}{\rho N} \sum_{i\neq j}  \expval{ \partial_r \vb*r \cdot \grad_{\vb*r} \delta(\vb*r-\vb*r_{ij})}.
\end{equation}
Using $\partial_r {\vb*r} = \vb*r / r$ and exploiting symmetry between indices $i$ and $j$, we express this as:
\begin{equation}
\partial_r g(\vb*r) = \frac{1}{\rho N} \sum_{i\neq j} \expval{ \frac{\vb*r}{r} \cdot \frac{(\grad_{\vb*r_j} - \grad_{\vb*r_i})}{2} \delta(\vb*r-\vb*r_{ij})}\!.
\end{equation}
Splitting the expectation value into two contributions and focusing only on one of them, we integrate by part:
\begin{align}
&\expval{ \frac{\vb*r}{r} \cdot \frac{1}{2}\grad_{\vb*r_j}  \delta(\vb*r-\vb*r_{ij})} = \\
&= \int \frac{\dd^N \vb*r}{Z} e^{-\beta U} \frac{\vb*r}{r} \cdot \frac{1}{2}\grad_{\vb*r_j} \delta(\vb*r-\vb*r_{ij}) \\
&= - \int \frac{\dd^N \vb*r}{Z} \frac{1}{2}\grad_{\vb*r_j} e^{-\beta U} \cdot \frac{\vb*r}{r}   \delta (\vb*r-\vb*r_{ij})\\
&= -\int \frac{\dd^N \vb*r}{Z}  \frac{\beta}{2}  \vb f_j  e^{-\beta U}  \cdot  \frac{\vb*r}{r}\delta(\vb*r-\vb*r_{ij}) ,
\end{align}
where $\vb f_j = -\grad_{\vb*r_j} U$ is the total force acting on particle $j$. 
The boundary term is zero assuming that the Boltzmann weight (and thus the pair potential) decays sufficiently fast. A similar expression holds for the second term with $\grad_{\vb*r_i}$. Combining both contributions, we find:
\begin{equation}
\partial_r g(\vb*r) = \frac{1}{\rho N} \sum_{i\neq j} \expval{ \beta \frac{(\vb f_i - \vb f_j)}{2} \cdot \frac{\vb*r}{r} \delta(\vb*r-\vb*r_{ij}) }.
\end{equation}
For homogeneous, isotropic systems, the RDF depends only on the radial distance $r = |\vb*r|$. We extract this radial dependence by averaging over solid angles:
\begin{equation}
\int_{\mathbb S_{d-1}} \dd \vb\Omega_{\hat r} \partial_r g(\vb*r) = \Omega_d \partial_r g(r) = \Omega_d g'(r),
\end{equation}
where $\Omega_d = 2\pi^{d/2} / \Gamma (d/2)$ is the surface of the $d$-dimensional unit sphere. At this point, the only term which depends on the angular components is the Dirac distribution, that satisfies the identity:
\begin{equation}
\int_{\mathbb S_{d-1}} \dd \vb\Omega_{\hat r} \ \delta(\vb*r-\vb*r_{ij}) = \frac{1}{r^{d-1}} \delta(r - r_{ij}).
\end{equation}
Altogether, using the symmetry between $i$ and $j$ to restrict the sum we obtain:
\begin{equation}
g'(r) = \frac{1}{\rho N} \expval{\sum_{i <j} \beta ({\vb f_i - \vb f_j}) \cdot \frac{\vb*r_{ij}}{\Omega_d r_{ij}^{d}} \delta(r - r_{ij}) }.
\end{equation}
We can then integrate from $0$ to $r$ and noting that $g(0) = 0$ for quasi-hard core potentials we get:
\begin{equation}
g(r) = \frac{1}{\rho N}  \expval{\sum_{i <j} \beta ({\vb f_i - \vb f_j}) \cdot \frac{\vb*r_{ij}}{\Omega_d r_{ij}^{d}} \Theta(r - r_{ij})}.
\end{equation}
Alternatively, one can integrate $g'(r)$ from $r$ to $\infty$ and impose the condition $g(r \to \infty) = 1$, which holds for systems lacking long-range order:
\begin{equation}
g(r)\! = 1 -\frac{1}{\rho N}\! \expval{\sum_{i< j}\! \beta ({\vb f_i - \vb f_j})\! \cdot\! \frac{\vb*r_{ij}}{\Omega_d r_{ij}^{d}} \Theta(r_{ij}\! - r)\!}\!.
\end{equation}
These expressions are the two equivalent forms of the Borgis formula, expressing the RDF directly in terms of interparticle forces. These expressions differ from the original work \cite{borgisComputationPairDistribution2013} by a factor $2$ that was omitted, as already pointed out in \cite{purohitForcesamplingMethodsDensity2019}.

These expressions do not rely on a specific form of the potential energy $U(\{\vb*r_k\})$, other than its isotropy and the presence of a (quasi-)hard core, only on the existence of well-defined forces $\vb f_i$ acting on each particle. Thus, these relations hold for any system where forces can be defined, including those with many-body interactions or non-pairwise potentials.

\section{Justification of iteration formula}

\label{Apa}

To ensure the self-consistency of the manuscript, we summarize the key arguments from Schommers~\cite{schommersPairPotentialLiquid1973} that underpin the Iterative Boltzmann Inversion approach, and discuss their implications for the reliability of the algorithm. While the mathematical foundations of the inverse problem has been studied extensively~\cite{delbaryGeneralizedNewtonIteration2020}, a general and rigorous convergence proof for the IBI algorithm itself is lacking.

For simplicity, we work in dimensionless units, absorbing $\beta$ into the potential $U(r)$. Consider a system with a true pair interaction potential $U^{(0)}(r)$ and its corresponding pair correlation function $g^{(0)}(r)$. These are related via the cavity function $\gamma^{(0)}(r)$, defined as
\begin{equation}
g^{(0)}(r) = \gamma^{(0)}(r) e^{- U^{(0)}(r)},
\end{equation}
where $\gamma(r)$ captures all correlations beyond the potential of mean force. Now, suppose we have guessed a potential $U^{(1)}(r) \neq U^{(0)}(r)$, which yields a pair correlation function $g^{(1)}(r)$ through a corresponding $\gamma^{(1)}(r)$. By Henderson's uniqueness theorem, the resulting $g^{(1)} \neq g^{(0)}$. However, if the $\gamma$ functions are close, i.e. $\gamma^{(1)}(r) = \gamma^{(0)}(r) + \Delta\gamma(r)$ with $|\Delta\gamma(r)| \ll |\gamma^{(0)}(r)|$, then we can write
\begin{equation}
\gamma^{(1)}(r) = g^{(1)}(r)e^{ U^{(1)}(r)} = g^{(0)}(r)e^{ U^{(0)}(r)} + \Delta\gamma(r).
\end{equation}
Rearranging and taking logarithms gives
\begin{equation}
U^{(1)}(r)-U^{(0)}(r) = \log \left( \frac{g^{(0)}(r)}{g^{(1)}(r)} \right) + \log \left(1 + \frac{\Delta\gamma(r)}{\gamma^{(0)}(r)}\right).
\end{equation}
Expanding to first order leads to 
\begin{equation}
U^{(1)}(r)-U^{(0)}(r) \approx \log \left( \frac{g^{(0)}(r)}{g^{(1)}(r)} \right) + \mathcal{O}\left(\frac{\Delta\gamma}{\gamma^{(0)}}\right).
\end{equation}
This relation provides two useful insights. Firstly, it connects the difference between the true and guessed potentials to the ratio of their pair correlation functions, motivating the IBI update rule Eq.~(\ref{eq: IBI-Schommers scheme}). Secondly, the error in this estimate is controlled by the relative deviation $\Delta\gamma/\gamma^{(0)}$, rather than by the  difference between the potentials themselves. This means that convergence hinges on the similarity of the many-body environments between iterations, not on an initially accurate guess for $U(r)$. 

The condition of closeness in $\gamma$, rather than in $U$, is significantly weaker and more practical. For example, in the low-density limit, the virial expansion gives $\gamma(r) = 1 + \mathcal{O}(\phi)$, where $\phi$ is the packing fraction. Therefore, for dilute systems, $\gamma(r)$ is close to unity regardless of the details of the potential, making the iterative correction reliable. This explains why initializing IBI with $U(r)=0$ often works for dilute systems~\cite{stonesModelFreeMeasurementPair2019}.

For denser systems, one may worry that strong many-body correlations could cause $\gamma(r)$ to differ significantly between iterations, potentially hindering or destabilizing convergence. To address this concern, we recall the arguments by Soper~\cite{soperEmpiricalPotentialMonte1996}, who addressed the question of convergence at arbitrary density. First of all, using the notation $\Delta U = U^{(1)}-U^{(0)}$ and $\Delta g = g^{(1)}-g^{(0)}$, Henderson's theorem provides a global constraint:
\begin{equation}
\int \Delta U(r) \Delta g(r) \, d^dr < 0,
\end{equation}
where the integral is over all space. This inequality indicates that, on average, it is possible to predict the sign of the variation in the pair correlation function given a change in the potential, and vice versa. The relative variation of $g$ and $U$ is at the core of the iteration rule in IBI.   

To understand it better, one can decompose $g(r)$ as $g(r) = g^{(p)}(r) + g^{(m)}(r)$, where $g^{(p)}(r)=e^{- U(r)}$ is the dilute contribution and $g^{(m)}(r)$ captures many-body effects via the potential of mean force $W(r)$: $g^{(m)}(r)=e^{-W(r)} - e^{-U(r)}$. It is easy to show that, for all $r$, $\Delta U(r) \Delta g^{(p)}(r) < 0$, so that
\begin{equation}
\int \Delta U(r) \Delta g^{(p)}(r) \,d^dr < 0.
\end{equation}
Combining this with the previous equation yields a bound on the many-body contribution:
\begin{equation}
\int \Delta U(r) \Delta g^{(m)}(r) \,d^dr < - \int \Delta U(r) \Delta g^{(p)}(r) \,d^3r.
\end{equation}
This shows that the relative variations of the many-body part $\Delta g^{(m)}$ and that of $\Delta U$ at a given $r$ are not directly related. 
Yet, the variation of the many-body contribution $g^{(m)}$ with $\Delta U$ is bounded by a positive number controlled by the dilute limit. For this reason, the use of the potential of the mean force in Schommers update rule in Eq.~(\ref{eq: IBI-Schommers scheme}) is justifiable even at finite density, although a rigorous convergence proof remains elusive.

\section{Stability analysis}

\label{Apb}

Even in the absence of a convergence proof, one can show that the iteration is stable. Namely, one can show that a small perturbation of the potential near the correct fixed point disappears by iterating Eq.~(\ref{eq: IBI-Schommers scheme}), rather than explodes. To establish stability, we adopt a continuous‑time view of the iteration. The iterative update can then be rewritten as
\begin{equation}
    \dv{U_t(r)}{t}=
    \alpha\,
    \log  \Bigl(\frac{g_t(r)}{g_{\rm ref}(r)}\Bigr).
\end{equation}
Here $t$ is now a real variable, and the positive constant $\alpha$ (often introduced for numerical convergence) does not affect the stability analysis. The radial distribution function $g_t(r)$ is a functional of the current potential $U_t$,
\begin{equation}
    g[U](r)=e^{-W[U](r)}=e^{-U(r)}\gamma[U](r),
\end{equation}
where $W[U](r)$ is the potential of mean force and $\gamma[U](r)$ the cavity distribution function~\cite{hansenTheorySimpleLiquids2013} associated to $g[U](r)$.

Now consider a potential close to the target $U_{\text{ref}}(r)$:
\begin{equation}
U(r)=U_{\text{ref}}(r)+\delta U(r),\qquad
\frac{|\delta U|}{|U_{\text{ref}}|}\ll1.
\end{equation}
Since $g[U_{\text{ref}}](r)=g_{\text{ref}}(r)$, the evolution of the perturbation is
\begin{equation}
    \begin{split}
    \dv{\delta U(r)}{t}
    &=\alpha\,
    \log\!\Bigl(\frac{g[U_{\text{ref}}+\delta U](r)}
    {g_{\text{ref}}(r)}\Bigr)\\
    &=\alpha\,
    \log\!\Bigl(
    \frac{e^{-U_{\text{ref}}-\delta U}\,
    \gamma[U_{\text{ref}}+\delta U](r)}
    {e^{-U_{\text{ref}}}\,\gamma[U_{\text{ref}}](r)}
    \Bigr)\\    
    &=-\alpha\,\delta U(r)
    +\alpha\,
    \log\!\Bigl(
    \frac{\gamma[U_{\text{ref}}+\delta U](r)}
    {\gamma[U_{\text{ref}}](r)}\Bigr) +O(\delta U^{2}).
\end{split}
\end{equation}
Expanding $\gamma$ to first order in $\delta U$, we get 
\begin{equation}
    \dv{\delta U(r)}{t}  = -\alpha\,\delta U
    +\alpha\,
    \frac{1}{\gamma[U_{\text{ref}}]}
    \eval{\frac{\delta\gamma}{\delta U}}_{U_{\text{ref}}}\!\delta U
    +O(\delta U^{2}).    
\end{equation}
In the low‑density limit $\gamma[U](r)\approx1$ for all potentials; hence $\delta\gamma/\delta U\approx0$ and we have
\begin{equation}
    \dv{t}\delta U\simeq-\alpha\,\delta U,
\end{equation}
giving
\begin{equation}
    \delta U(r)=\delta U_{0}(r)\,e^{-\alpha t}.
\end{equation}
Thus, at low density, any sufficiently small initial perturbation decays exponentially, confirming stability of the iteration near the fixed point. As for the convergence formula above, it is difficult to extend this argument beyond the dilute limit, which thus simply serves as a guide for denser systems.

\end{document}